\documentclass[%
 reprint,
 amsmath,amssymb,
 aps,
floatfix,
]{revtex4-1}

\usepackage{graphicx}% Include figure files
\usepackage{dcolumn}% Align table columns on decimal point
\usepackage{bm}% bold math
\usepackage{hyperref}% add hypertext capabilities
 \usepackage[compat=1.1.0]{tikz-feynman}
 \usepackage{physics}

\begin{document}

%\preprint{APS/123-QED}

\title{Non-relativistic limit of the interaction of massive neutrinos}% Force line breaks with \\

\author{Andre Fabiano Steklain}
 \affiliation{Mathematics Department, Federal University of Technology - Paran\'a.}%Lines break automatically or can be forced with \\


\date{\today}% It is always \today, today,

\begin{abstract}
In this work we consider the electroweak interaction between two massive neutrinos in the non-relativistic limit. We find that in this limit the interaction can be null.
\end{abstract}

%\pacs{03.70.+k,12.15.−y,13.15.+g,14.60.Pq,14.70.Hp}
%Use showkeys class option if keyword
\maketitle

%\tableofcontents

\section{Introduction}

In Standard Model the neutrino is considered to be a massless particle \cite{Zuber2011}. The solar neutrino problem and its solution by considering neutrino oscillations, however, imply that neutrinos must be massive, although the masses are very small. On the contrary of massless particles, that always propagate with the speed of light $c$, massive particles can be considered in a rest frame, or in a non-relativistic limit, where velocities are small in comparison with $c$. 

The Coulomb potential can be obtained from the quantum electrodynamics in the non-relativistic limit \citep{PeskinSchroeder1995}. The repulsion between fermions with same charge is explained by the vector nature of the exchanged particle, in this case, the photon. Similar arguments provide that when a scalar or tensor particle is exchanged, the force between them is universally atractive.

Neutrinos have no charge, so they only interact through weak interaction or gravity. As the masses are small, the gravitation interaction can be disregarded when compared to the weak force. The weak interaction is mediated by vector bosons, namely, $W^{\pm}$ and $Z^0$. A na\"{\i}ve comparison with the electrodynamic interaction can lead to the conclusion that a the weak force between two neutrinos is also repulsive, as pointed out in works that consider neutrino superfluids \cite{Kapusta2004}. However, the non-relativistic limit of the weak interaction must be carried out in detail. The aim of this work is to provide such calculation for the non-relativistic limit of the weak interaction between two massive neutrinos.

\section{Calculation}

\begin{figure}[h!]
    \centering
% \feynmandiagram [horizontal=a to b] {
%  a  -- [fermion,edge label'=\(p_1'\)] i1, i2 -- [fermion,edge label'=\(p_1\)] a, 
%  a -- [charged scalar, edge label=\(Z_0\)] b,
%  f1 -- [fermion,edge label=\(p_2\)] b -- [fermion,edge label=\(p_2'\)] f2,};
\includegraphics[scale=0.08]{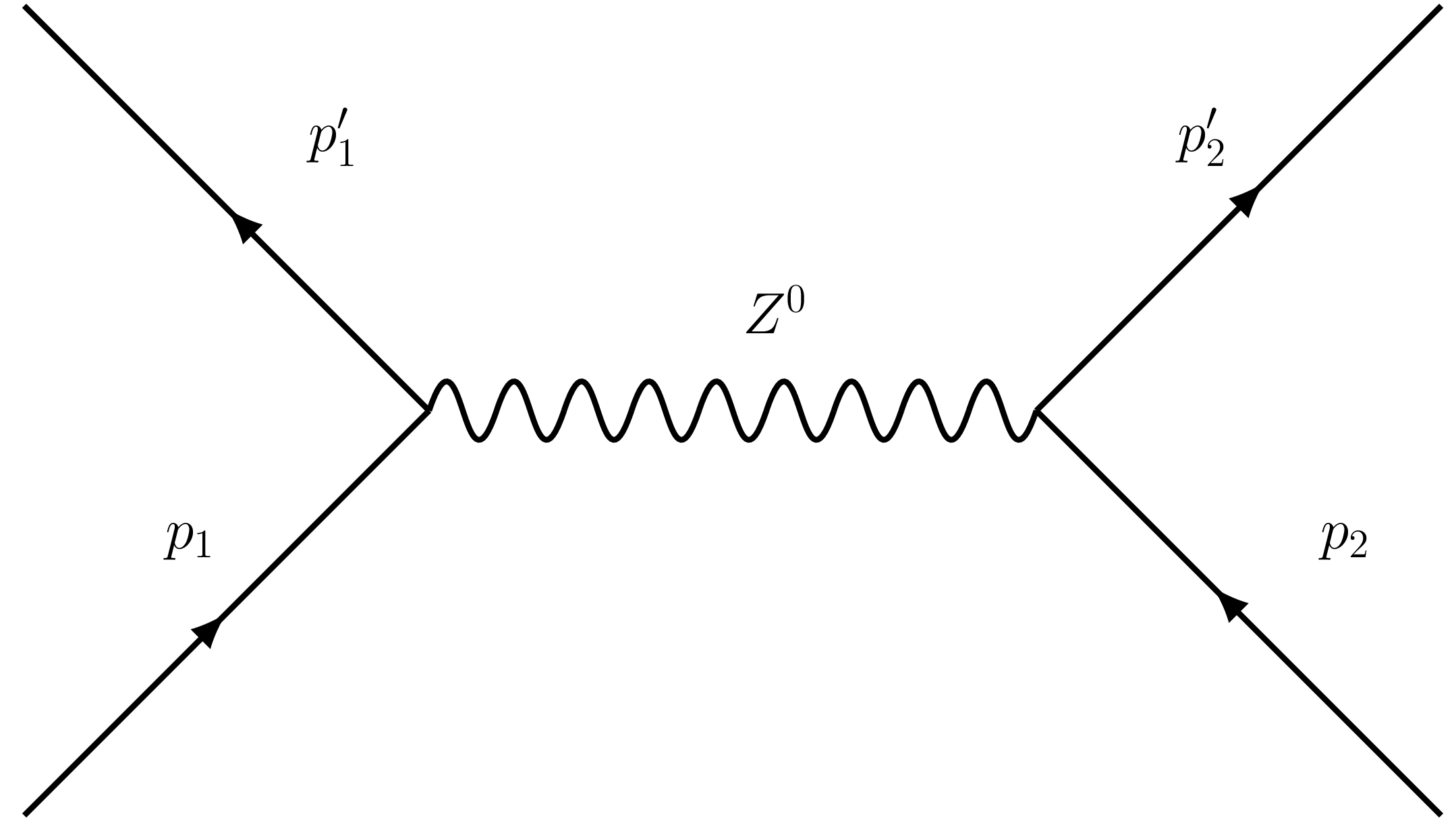}
    \caption{Feynman diagram of the weak interaction between two distinguishable neutrinos.}
    \label{fig:diagram}
\end{figure}

In this work we consider the scattering of two massive, distinguishable neutrinos. If gravity is not taken into account, the interaction between the two neutrinos occurs through the exchange of the massive neutral vector meson $Z^0$. In this case, only the diagram in Figure \ref{fig:diagram} is allowed \cite{Flowers1976}. 

According to the Feynman rules the matrix element $\mathcal{M}$ is given by \citep{Flowers1976}

\begin{widetext}
\begin{equation}
    i\mathcal{M}= (ig)^2 \bar{u}(p_1')\gamma ^\mu (1- \gamma ^5)u(p_1)\left[\frac{-ig_{\mu \nu}}{q^2 - M_Z^2}\right]\bar{u}(p_2')\gamma ^\mu (1- \gamma ^5)u(p_2),
\end{equation}
\end{widetext}
in the Feynman gauge. In this equation $g$ is the $SU(2)_L$ coupling constant \cite{aitchison2012gauge}, $M_Z$ s the mass of $Z^0$ and $q^\mu = p^\mu - p'^\mu$. We consider the metric tensor with signature $(+,-,-,-)$ and take the gamma matrices in the chiral (Weyl) representation \cite{PeskinSchroeder1995}.  

We now proceed to the calculation of the terms $\bar{u}(p')\gamma ^\mu (1- \gamma ^5)u(p)$ in the non-relativistic limit by the use of an argument similar to that found in \cite{PeskinSchroeder1995}. The well-known Gordon identity,
\begin{equation}
    \bar{u}(p)\gamma ^\mu u(p) = \bar{u}(p) \left[ \frac{p'^\mu+p^\mu}{2m} +i\frac{\sigma ^{\mu \nu}q_\nu}{2m}\right] u(p)
\end{equation}
can be easily modified in order to accomodate the extra $\gamma ^5$ term (the derivation of the modified Gordon identity can be found in Appendix \ref{ap:modifiedgordon}):

\begin{widetext}
\begin{equation}
\label{gordonmodified}
    \bar{u}(p)\gamma ^\mu (1-\gamma ^5) u(p) = \bar{u}(p) \left[ \frac{p'^\mu+p^\mu}{2m} -i\frac{\gamma ^5 \sigma ^{\mu \nu}(p'_\nu+p_\nu)}{2m}+i\frac{\sigma ^{\mu \nu}q_\nu}{2m}-\frac{q^\mu \gamma ^5}{2m}\right] u(p).
\end{equation}
\end{widetext}

In the non-relativistic approximation we shall keep terms only to lowest order in the 3-momenta. Thus, up to $\mathcal{O}({\bf p^2,p'^2, ...})$ we have $p = (m,\mathbf{p})$ and $p' = (m,\mathbf{p'})$. Using these expressions, we have
\begin{align}
    (p-p')^2=-|\mathbf{p}-\mathbf{p'}|+\mathcal{O}(\mathbf{p}^4);\\
    u(p) = \sqrt{m} \left( \begin{array}{c} \xi  \\ \xi  \end{array} \right).
\end{align}
In this approximation we have, for the $\mu =0$ component,
\begin{equation}
    \bar{u}(p')\gamma ^0 \left(1-\gamma ^5 \right)u(p) \approx 2 m \xi ^\dagger \xi = 2m.
    \label{muoresult}
\end{equation}
For the electron-electron interaction in QED the other terms can be neglected when compared with the $\mu =0$ term. The presence of the $\gamma ^5$ term, however, makes the spatial terms of the metric relevant. Using the modified Gordon identity we have for $\mu=i$ components
\begin{equation}
     \bar{u}(p')\gamma ^i \left(1-\gamma ^5 \right)u(p) \approx u^\dagger (p')\gamma ^5 \gamma ^i u(p).
     \label{mui}
\end{equation}
In the chiral representation the gamma matrices are given by \cite{PeskinSchroeder1995}:
\begin{align}
    \gamma ^0 = \left(\begin{array}{cc}
        0 & 1 \\
        1 & 0
    \end{array} \right),\\
    \gamma ^i = \left(\begin{array}{cc}
        0 & \sigma ^i \\
        -\sigma ^i & 0
    \end{array} \right),\\
    \gamma ^5 = \left(\begin{array}{cc}
        -1 & 0 \\
        0 & 1
    \end{array} \right),
\end{align}
where $\sigma ^i$ are the Pauli matrices. Using this representation in equation (\ref{mui}) we have
\begin{equation}
      \bar{u}(p')\gamma ^i \left(1-\gamma ^5 \right)u(p) \approx -2m \xi ^\dagger \sigma ^i \xi.
      \label{muiresult}
\end{equation}
By combining equations (\ref{muoresult}) and (\ref{muiresult}) we obtain, for the $\mathcal{M}$ matrix:
%\begin{equation}
%    i \mathcal{M}= i g^2 \frac{4m^2}{q^2-M_Z ^2} \left[1- \sum _{i=1} ^3 \left( \xi ^\dagger \sigma ^i \xi \right)^2 \right].
%\end{equation}
\begin{widetext}
\begin{equation}
%    i \mathcal{M}= i g^2 \frac{4m^2}{q^2-M_Z ^2} \left[\left( \xi ' _1 {}^\dagger  \xi _1 \right)\left( \xi '  _2 {}^{\dagger}  \xi _2 \right)- \sum _{i=1} ^3 \left( \xi ' _1 {}^\dagger \sigma ^i \xi _1 \right)\left( \xi '  _2 {}^\dagger \sigma ^i \xi _2 \right) \right].
i \mathcal{M}= i g^2 \frac{4m^2}{q^2-M_Z ^2} \left[1 - \sum _{i=1} ^3 \left( \xi _1 {}^\dagger \sigma ^i \xi _1 \right)\left( \xi  _2 {}^\dagger \sigma ^i \xi _2 \right) \right].
\end{equation}
\end{widetext}

From this expression we note that, in addition with the first term, that provides a repulsive potential, there are four other terms with inverted signal, that provide an attractive counterpart. Consider, for instance, that the spin of both particles are aligned in the $z$ direction. It is easy to see that in this case the matrix element is zero. In fact, it is possible to demonstrate that this matrix is null for any spin state. Following the procedure by Peskin and Schroeder we compare this result with the Born approximation to the scattering amplitude in non-relativistic quantum mechanics, given in terms of the potential function $V(\mathbf{x})$ by \citep{PeskinSchroeder1995}
\begin{equation}
\bra{\mathbf{p'}}iT\ket{\mathbf{p}}=-i\tilde{V}(\mathbf{q})(2\pi)\delta (E_\mathbf{p'}-E\mathbf{p}),
\end{equation}
where $\tilde{V}(\mathbf{q})$ is the Fourier transform of the potential $V(\mathbf{x})$ and $\mathbf{q}=\mathbf{p'}-\mathbf{p}$. In this case we find that the potential is $V(\mathbf{x})=0$.

This result can be understood if we interpret the weak interaction in the $V-A$ theory where, in addition to the vector current ($V$), there is an axial vector current ($A$). The vector current generate a repulsive interaction between the two fermions, but the axial vector current provide an attractive component in the non-relativistic limit. For neutrino interactions this two components cancel each other exactly, in a way that the total interaction is zero. 

For weak interactions regarding other types of fermions it is necessary to consider the dependency of the coupling factor for the $Z^0$ current on $c_V$ and $c_A$, and charged interactions. From the complete analysis of these interactions we can obtain the nature of the weak force in the non-relativistic limit.

\section{Acknowledgements}

The author thanks Edgard Amorim for all the useful discussions.

\bibliography{references}

\appendix

\section{Derivation of the Modified Gordon Identity}
\label{ap:modifiedgordon}

We will consider the term $ \bar{u}(p')\gamma ^\mu \gamma ^5 u(p)$. Using the Dirac equation in the momentum space
\begin{equation}
(\gamma ^\mu p _\mu -m)u(p)=0, 
\end{equation}
we have the following relation:
\begin{align}
\bar{u}(p')\gamma ^\mu \gamma ^5 u(p) &= \bar{u}(p')\gamma ^\mu \gamma ^5 \gamma ^\mu \frac{p _\mu}{m} u(p) \\ &= \bar{u}(p')\left[-\gamma ^5 \frac{p _\mu}{m} +i\gamma ^5 \sigma ^{\mu \nu} \frac{p_\nu}{m}  \right]u(p),
\end{align}
using the anticommutation properties of the gamma matrices and the spin matrices given by
\begin{equation}
\sigma ^{\mu \nu} = \frac{i}{2}\left[\gamma ^\mu, \gamma ^\nu\right]
\end{equation}
similarly, using the conjugate counterpart of Dirac equation:
\begin{equation}
\bar{u}(p')(\gamma ^\mu p' _\mu -m)=0, 
\end{equation}
we have
\begin{align}
\bar{u}(p')\gamma ^\mu \gamma ^5 u(p) = \bar{u}(p')\left[\gamma ^5 \frac{p' _\mu}{m} -i\gamma ^5 \sigma ^{\nu \mu} \frac{p'_\nu}{m}  \right]u(p).
\end{align}
Combining these two results we have, finally, the identity:
\begin{widetext}
\begin{align}
\bar{u}(p')\gamma ^\mu \gamma ^5 u(p) = \bar{u}(p')\left[\gamma ^5 \frac{(p'-p) _\mu}{2m} +i\gamma ^5 \sigma ^{\mu \nu} \frac{(p+p')_\nu}{2m}  \right]u(p).
\end{align}
\end{widetext}
When combined with the original Gordon identity we obtain finally the equation \ref{gordonmodified}.

\end{document}